\documentclass[conference]{IEEEtran}
\IEEEoverridecommandlockouts
\usepackage{cite}
\usepackage{amsmath,amssymb,amsfonts}
\usepackage{algorithmic}
\usepackage{graphicx}
\usepackage{textcomp}
\usepackage{xcolor}
\usepackage{multirow}
\usepackage{bbm}
\usepackage[a4paper, total={184mm,239mm}]{geometry}
\usepackage{comment}

\usepackage{enumitem}
\setlist[itemize]{leftmargin=*}
\setlist[enumerate]{leftmargin=*}

\def\BibTeX{{\rm B\kern-.05em{\sc i\kern-.025em b}\kern-.08em
    T\kern-.1667em\lower.7ex\hbox{E}\kern-.125emX}}

\usepackage{hyperref}
\hypersetup{
    colorlinks=true,
    citecolor=blue,
    linkcolor=blue,
    urlcolor=black,
}
\usepackage{cleveref}

\usepackage{caption}
\newcommand{\pname}{\mbox{{VeriBug}}}

\newcommand{\eg}{\mbox{{\em e.g.}}}
\newcommand{\ie}{\mbox{{\em i.e.}}}
\newcommand{\cf}{\mbox{{c.f.}}}

\newcommand{\bem}[1]{{\bf\em #1}}
\newcommand{\goldmine}{{\scshape GoldMine}}

\newcommand{\addt}[1]{\textcolor{black}{#1}}

\newcount\Comments  
\newcommand{\kibitz}[2]{\ifnum\Comments=1{\color{#1}{#2}}\fi}

\begin{document}

\title{$\pname$: An Attention-based 
Framework for Bug-Localization in Hardware Designs
}

\author{%
\IEEEauthorblockN{Giuseppe~Stracquadanio$^{1,2}*$, Sourav~Medya$^1$, Stefano~Quer$^2$ and Debjit~Pal$^1$}
  \IEEEauthorblockA{$^1$\textit{University of Illinois Chicago, Chicago, IL, USA} \; $^2$\textit{Politecnico di Torino, Torino, Italy}}
  \IEEEauthorblockA{\{gstrac3, medya, dpal2\}@uic.edu, stefano.quer@polito.it}
}

\maketitle
\begingroup\renewcommand\thefootnote{*}
\footnotetext{Work done while Giuseppe was studying toward his Master's degree. This work was funded by a startup fund from UIC.}
\endgroup
\thispagestyle{plain}


\begin{abstract}
In recent years, there has been an exponential growth in the size and complexity of System-on-Chip designs targeting different specialized applications. The cost of an undetected bug in these systems is much higher than in traditional processor systems as it may imply the loss of property or life.
The problem is further exacerbated by the ever-shrinking time-to-market and ever-increasing demand to churn out billions of devices. Despite decades of research in simulation and formal methods for debugging and verification, 
it is still one of the most time-consuming and resource intensive processes in contemporary hardware design cycle. 
In this work, we propose $\pname$, which leverages recent advances in deep learning to accelerate debugging at the Register-Transfer Level and generates explanations of likely root causes. First, $\pname$ uses control-data flow graph of a hardware design and learns to execute design statements by analyzing the context of operands and their assignments. Then, it assigns an importance score to each operand in a design statement and uses that score for generating explanations for failures. Finally, $\pname$ produces a heatmap highlighting potential buggy source code portions. 
Our experiments show that $\pname$ can achieve an average bug localization coverage of $82.5\%$ on open-source designs and different types of injected bugs. 
\end{abstract}

\section{Introduction 
}
\label{sec:intro}

Simulation and formal verification are two complementary verification techniques.
Given a design property, formal verification proves the property holds for every point of the search space.
Simulation verifies the property by pseudo-randomly testing a small subset of the search space.
The main drawback of the former is its unscalability, whereas the obvious flaw of the second is its inability to prove that a property holds for every point of the search space.
With these limitations and the ever-increasing time-to-market and design complexity, developing new hardware and software verification methods is mandatory.

In this work, we specifically focus on bug localization in hardware designs.
Bug localization techniques relying on formal methods, such as Binary Decision Diagrams (BDDs)~\cite{bdd2}, Bounded Model Checking (BMC)~\cite{bmc}, Interpolants~\cite{interpolants}, IC3~\cite{ic3},
and other SAT-based methods offer a systematic and rigorous approach to identifying and localizing bugs in hardware designs. However, they are computationally intensive, require additional expertise, and imply conspicuous effort for specifying complex mathematical logic models. At the same time, bug localization in simulation-based workflows is addressed by collecting failure traces and identifying common patterns~\cite{pal2016vlsid}.
Patterns corresponding to common paths in failure runs can be mapped to the original source code to identify suspicious code zones that require further inspection. A critical problem is that the localized suspicious code zones lack explanation
and may provide not much information about the actual bug.
Consequently, it is crucial to develop smarter techniques to reveal problems and localize them.

To reduce the gap between standard verification and everyday simulation techniques, we propose $\pname$, an automated bug-localization framework that harnesses the power of a data-driven approach and machine learning (ML). 
ML techniques for bug detection and localization have been studied in the software domain. However, we identify two main drawbacks of these approaches: (i) They directly extract features from specific code~\cite{bugram} and do not guarantee generalization to unseen code structures, and (ii) They approach the problem as a \textit{classification} task, where an entire program is classified as buggy or not buggy~\cite{bug_detection_dl}, based on features extracted from a \textit{static} analysis. These approaches suffer from poor generalization and require building a dataset with many programs containing one or more bugs. As such datasets are not readily available, this task becomes challenging.

%
These two limitations can also be transposed to the hardware domain. To overcome them, $\pname$ does not rely on 
code characteristics, but automatically learns features from lower abstraction levels, such as Abstract Syntax Trees (ASTs). These learned features are 
\textit{design-agnostic}, allowing $\pname$ to generalize on unseen designs without any retraining. We 
learn execution semantics
with a novel deep-learning architecture and use it to localize the root cause of a bug 
via comparison of learned semantics for failure and correct simulation traces.
$\pname$ does not need to be trained on a labeled design corpus 
as we train it on a \textit{proxy}-task to learn execution semantics directly from simulation traces. Our approach is 
fully integratable with current verification workflows without requiring additional time or artifacts. 

Our primary contributions are the followings: (i) $\pname$ represents the first approach to bug localization in hardware designs that automatically learns task-relevant features in a 
inductive and generalizable setting, (ii) We learn execution semantics features using free supervision by simulation traces and we show how these features can be used for providing bug localization insights and producing a localization heatmap, (iii) We show learned knowledge is transferable and generalizable to unseen designs, 
and (iv) We conduct a bug injection campaign on 4 different real designs, obtaining an average coverage of $82.5\%$ and localizing 85 injected bugs over a total of 103 observable bugs.


\section{Preliminaries}
\label{sec:prelim}


\noindent {\bf A Static Analyzer for Hardware Designs}: \goldmine~is an open-source hardware design analysis framework~\cite{pal2020goldmine}. It takes the design source code as input and performs light-weight source code analysis. Moreover, \goldmine~produces multiple artifacts, a few of which are: (i) Control-data flow graph (CDFG), which captures control flow and data flow among design variables, (ii) Variable dependency graph (VDG), which summarizes the control and data dependencies among design variables by abstracting operation details among design variables, and (iii) Cone-of-influence (COI) which captures the temporal relations among design variables when a design is unrolled for $n$ cycles. 
In this work, we have extensively used \goldmine-generated CDFG, VDG, and testbenches for developing $\pname$.

\smallskip 

\noindent {\bf Attention Networks}: 
The attention mechanism has been originally 
developed to improve the performances of encoder-decoder systems for the
machine translation tasks~\cite{bahdanau2016neural}. It weighs the relevance of each word in the input sequence when generating the next word of the output sequence. The attention mechanism generally allows neural networks to focus only on the important parts of the input data. 
We use 
dot-product attention to learn 
operand 
influence in design execution. 

\smallskip

\noindent {\bf Recurrent Neural Networks}:
Recurrent Neural Networks (RNNs) \cite{rnn} are a class of Artificial Neural Networks (ANN) 
tailored to process sequential data. RNNs 
maintain an \textit{internal state} or \textit{memory} of previous inputs. This inherent memory makes them particularly suitable for scenarios where past information influences 
decision-making. 
Besides being used to make predictions on sequential data, RNNs exhibit remarkable effectiveness 
in \textit{embedding} variable-length sequential data into a compact, fixed-size embedding vector. 
In this work, we use a sophisticated RNN model, Long-Short Term Memory (LSTM), to generate 
representations of variable-length node paths 
in Abstract Syntax Trees (ASTs) of design source code.  



\section{Bug localization as Learning Problem}


We aim to root-cause a design failure and localize it to a subset of the design source code. {\em Given a Verilog hardware design ${\bf D}$ with ${\bf I}$ inputs, ${\bf O}$ outputs, and an observed failure $f$ at output ${\bf t} \in {\bf O}$, $\pname$ localizes the failure to a subset of likely suspicious source code fragments ${\cal S} \in {\bf D}$ and generates quantitative explanations of suspiciousness as such. 
}

\smallskip

\begin{figure}
    \centering
    \includegraphics[scale=0.30]{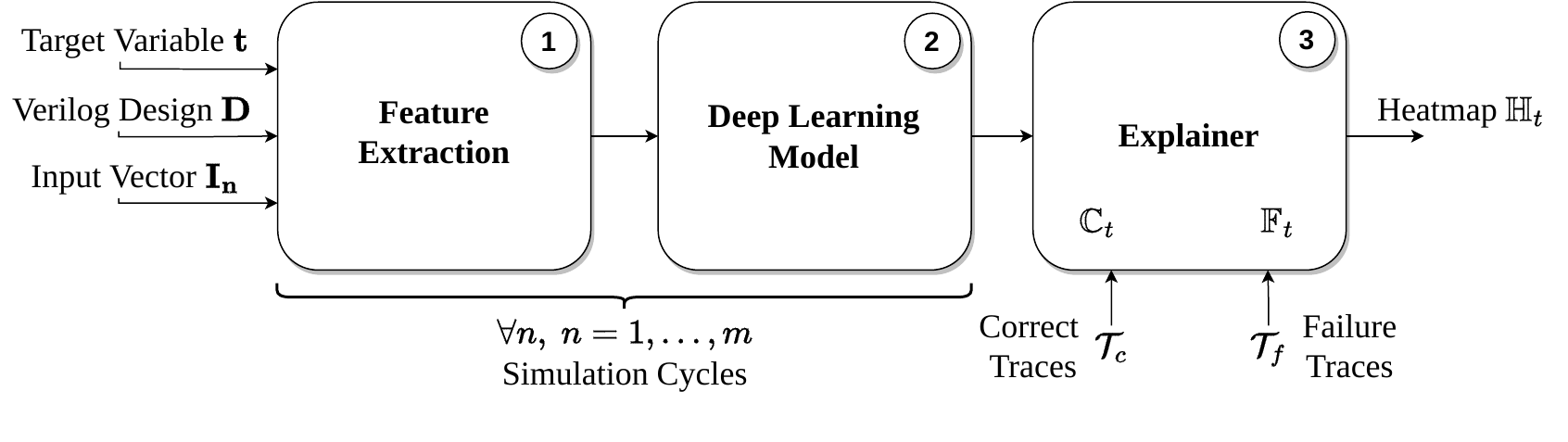}
    \vspace{-1mm}
    \caption{{\bf $\pname$ workflow.} (1) The Feature Extraction component extracts 
    \textit{features} from the 
    dynamic analysis of the
    design. (2) The Deep-Learning Model learns execution semantics 
    using these 
    features to predict 
    target values. (3) The Explainer component aggregates trace-level semantics into condensed execution information, producing  
    final heatmap $\mathbb{H}_t.$
    \label{fig:workflow}}
    \vspace{-7mm}
\end{figure}

\noindent {\bf Workflow and insights}:~\Cref{fig:workflow} shows the complete workflow of $\pname$. Given a design ${\bf D}$, an input vector ${\bf I}_n$, a set of correct simulation traces ${\cal T}_c$, and set of failure simulation traces ${\cal T}_f$, we propose to learn Verilog 
execution semantics automatically and use the learned knowledge 
to assign \textit{importance scores} to design statement operands. 
An insight is that 
these importance weights, computed using the concept of {\em attention}, contain critical design execution information that can be leveraged to {\em generate explanations for design failures}. After that, we aggregate such importance weights to concisely represent the design execution in passing failure traces. 
The aggregation results in two different 
maps, $\mathbb{F}_t$ and $\mathbb{C}_t$, respectively, for 
failing and correct traces. Then, we compute a \textit{suspiciousness score} for each design statement $l_k \in {\bf D}$ 
as ${\bf d}(\mathbb{F}_t (l_k) , \mathbb{C}_t (l_k))$, where ${\bf d}$ is a normalized norm-1 distance and $\mathbb{F}_t (\cdot)$ ($\mathbb{C}_t (\cdot)$) are the aggregated importance scores for a statement $l_k$ in the aggregated map 
$\mathbb{F}_t$ ($\mathbb{C}_t$). When the \textit{suspiciousness score} for $l_k$ is higher than a threshold,  we store $\mathbb{F}_t (l_k)$ importance scores in a final heatmap $\mathbb{H}_t$. Eventually, these score can be mapped back to RTL code to provide further visual guidance in locating the root cause. 
\section{Proposed methodology: $\pname$}
\label{sec:method}

\subsection{Overview of $\pname$ architecture}
\label{sec:workflow}


$\pname$ introduces a novel bug-localization technique powered by a \textit{deep learning} (DL) architecture. Its architecture consists of three key components: (i) Feature extraction, (ii) Deep learning model, and (iii) Explanation generation. We have summarized our proposed framework in~\Cref{fig:workflow}.

\subsection{Feature extraction}
\begin{figure}
\vspace{-3mm}
    \centering
    \includegraphics[scale=0.30]{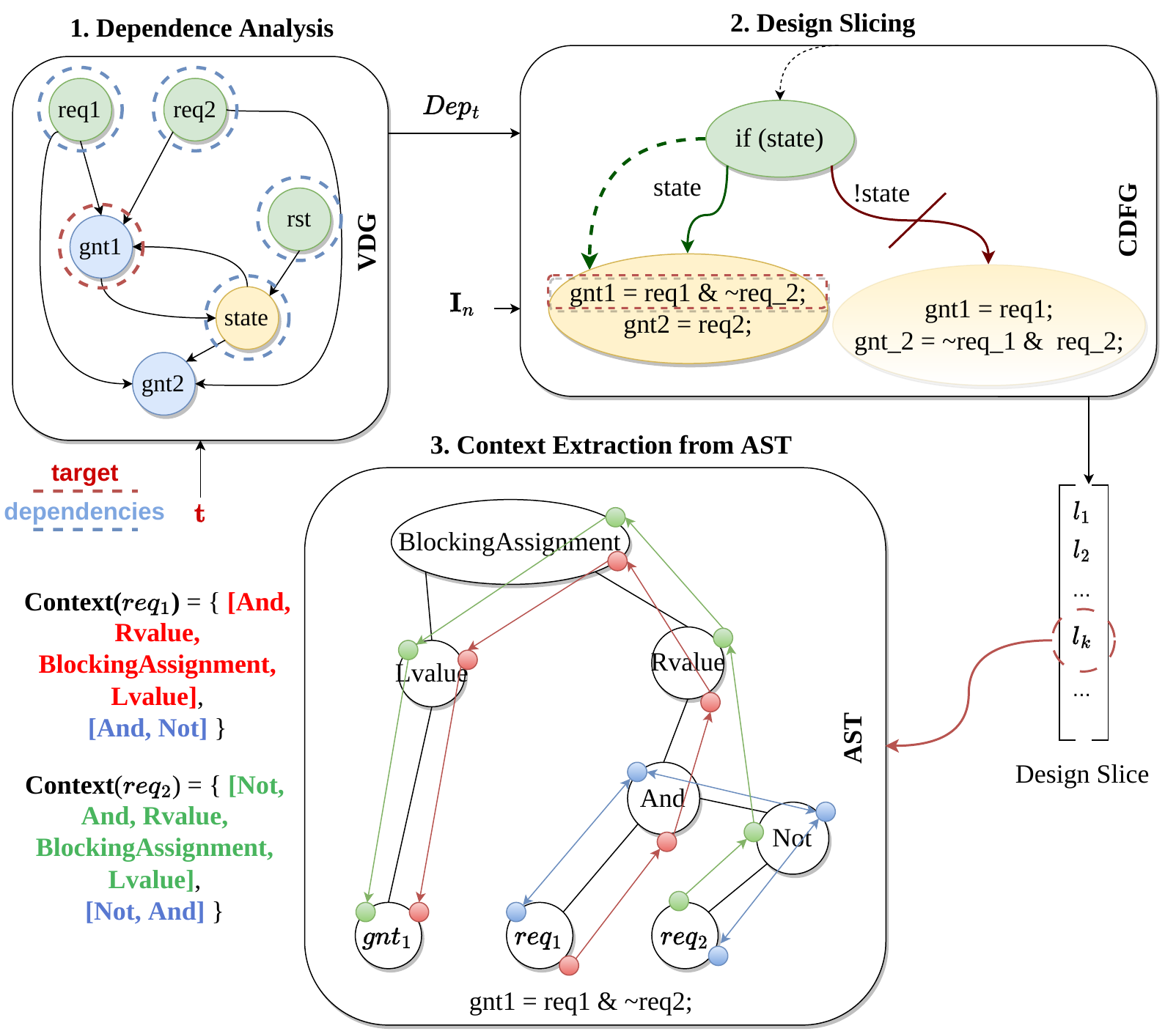}
    \caption{{\bf Feature Extraction Module}. (1) A dependence analysis produces the complete set $\mathbf{Dep}_t$ of control and data dependencies for the target output variable $\mathbf{t}$. (2) The slicing criterion uses the extracted set of dependencies $\mathbf{Dep}_t$ and the input vector $\mathbf{I}_n$ ($\eg\ \{req1 = 0, req2 = 1, ...\}$) to \textit{dynamically} slice the input design $\mathbf{D}$. (3) Context extraction is finally achieved for the operands of each slice statement, after encoding them to Abstract Syntax Trees (ASTs) In the figure, the context of $req_1$ is the list of paths $\{[And, Rvalue, BlockingAssignment,Lvalue],[And,Not]\}$. 
    \label{fig:feat_ext_preprocessing}}
    \vspace{-10mm}
\end{figure}

The first component 
\textit{Feature extraction} 
extracts knowledge ($\ie$, features) from the input design $\mathbf{D}$. 
\Cref{fig:feat_ext_preprocessing} presents an illustration of this entire module consisting three steps. We use 
\goldmine~to extract a Control Data Flow Graph (CDFG) and a Variable Dependency Graph (VDG) 
of the 
Verilog design $\mathbf{D}$, and to generate a {\em testbench} to simulate the design. 

\smallskip


\noindent {\bf Dependence analysis}:
We perform a \textit{dependence analysis} on the VDG 
to identify the set $\mathbf{Dep_t}$ of the \textit{control} and \textit{data} dependencies for $\mathbf{t}$, as shown in \Cref{fig:feat_ext_preprocessing}(1). This set is computed by reversing edges in the VDG and by starting a \textit{Depth First Search} (DFS) from the target node $\mathbf{t}$.  All the nodes that can be visited through a DFS are added to $\mathbf{Dep_t}$.

\smallskip

\noindent{\bf Design slicing
}: 
We use target variable $\mathbf{t}$ on the CDFG to extract design slices. Our slicing criterion simply includes a statement in the slice if the left-hand side (LHS) variable is in $\mathbf{Dep_t}$. The slicing criterion is directly applied on complete CDFG of a design. 
Furthermore, we exclude program slices involving branches 
which cannot be executed by a given input vector $\mathbf{I}_n$. Intuitively, if a statement is not \textit{executed} 
by $\mathbf{I}_n$, 
then it is certainly 
not the cause of a bug symptomatized at 
one of the 
outputs. After extracting all relevant statements, we translate them into 
Abstract Syntax Trees (ASTs) for further analysis. In~\Cref{fig:feat_ext_preprocessing}(1), we chose ${\bf gnt1}$ as $\mathbf{t}$ and use DFS to find ${\bf Dep_t}$ = \{${\bf req1, req2, state}$\}. We use ${\bf gnt1}$ and an input ${\bf I_n} = \{req1 = 0, req2 = 1, ...\}$ to extract the dynamic slice comprising the green and the yellow ellipses in~\Cref{fig:feat_ext_preprocessing}(2).~\Cref{fig:feat_ext_preprocessing}(3) shows the AST of one of the 
design statements $gnt1 = req1 \& \neg req2$ 
in the dynamic slice.

\smallskip

\noindent{\bf Context extraction from ASTs}:
The root node of an 
AST is the assignment type and the leaves are the input operands and the output variable. For the sake of brevity, we indicate the set of leaf nodes for the AST of $l_k$ as $\mathcal{L}(l_k)$. In~\Cref{fig:feat_ext_preprocessing}(3), the root node is of {\em BlockingAssignment} type whereas the  leaf nodes are variables, \eg, ${\bf req1}$, ${\bf req2}$, etc. We 
encode the relative structural information within an AST using 
\textit{leaf-to-leaf} paths. 
We extract all the paths for each input operand $op_i$ from the AST, \ie, 
$op_i \rightarrow op_k,\ \forall op_k \in \mathcal{L}(l_k) \setminus \{op_i\}$. In~\Cref{fig:feat_ext_preprocessing}(3), $req1 \rightarrow And \rightarrow Not \rightarrow req2$ is one such path. 
As this list represents the structural information for each operand, we refer it as the \textit{context of the operand} in a statement. 

\subsection{Our Deep-learning model}\label{sec:dl_model}
\begin{figure}
\vspace{-3mm}
    \centering
    \includegraphics[scale=0.29]{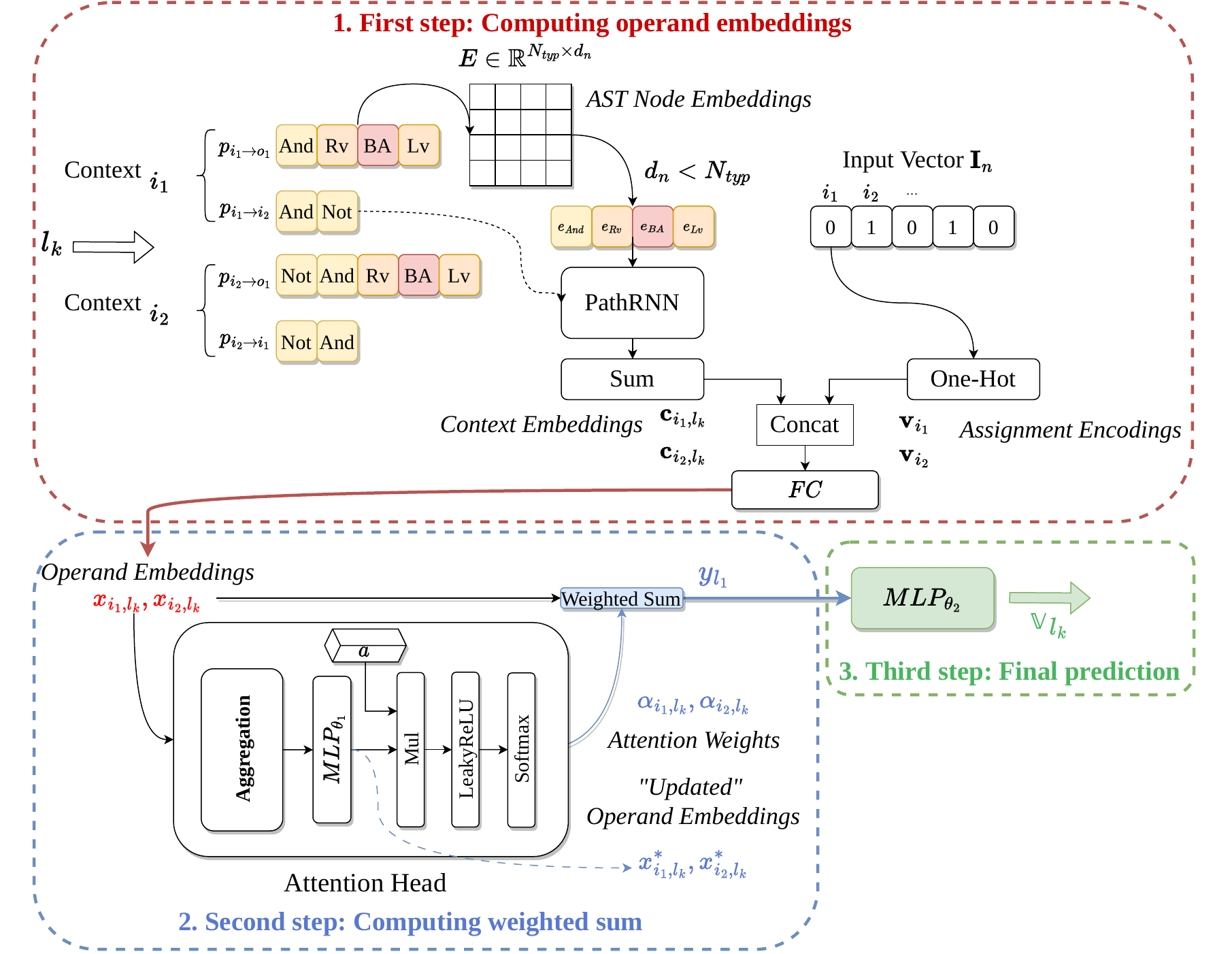}
    \caption{{\bf The Deep Learning Model of $\pname$.} 
    (1) For each statement $l_k$ and its AST, the model embeds operand contexts and encode their assignments to concatenate them in operand embeddings. (2) A weighted sum is computed with operand embeddings, using attention weights produced by our attention module. (3) A final prediction for $l_k$ output value is made from a statement-level embedding produced by weighted sum.  
    \label{fig:deep_learning_model}}
    \vspace{-6mm}
\end{figure}

The 
deep learning model leverages 
extracted knowledge 
from the feature extractor 
to {\em learn RTL execution semantics}. \Cref{fig:deep_learning_model} shows an overview of our model.  
We formulate the learning problem 
as \textendash~{\em predicting 
the output values of each executed statement 
using a combination of the input assignments and the context features as the input to our model}.
This formulation allows us to use signal values from the simulation traces as training data for 
supervision 
without requiring any \textit{labels}. 

\smallskip

\noindent {\bf Our learning task}:
For each 
statement $l_k$ sampled from the training set, the model aims to predict the output value assigned to the LHS of $l_k$, given 
input assignments in $\mathbf{I}_n$. 
$\pname$ learns to execute statements 
using operands' \textit{contexts} and their \textit{assignments}. To predict output values, 
$\pname$ assigns an attention weight ($\ie$, an importance score) to each operand in a statement. We 
use these attention weights for \textit{debugging} and \textit{bug localization} as they reveal 
\textit{explanation} about the execution. 


\smallskip

\noindent \bem{Operand embeddings}:
As mentioned above, $\pname$ learns to execute statements using 
the information about the operands. Thus, our model $\pname$ first generates expressive representations for each operand by embedding the contexts of operands from the AST and the assignments in two fixed-sized vectors. Specifically, for each path in a context (see an example of a context with multiple paths in Fig. \ref{fig:feat_ext_preprocessing}), a LSTM model, referred as {\em PathRNN} at~\Cref{fig:deep_learning_model}, is employed to produce a \textit{path embedding} $\mathbf{p_{i\rightarrow j}} \in \mathbb{R}^{d_c}$ from the paths consisting of sequence of nodes in the AST, where $i$ denotes the operand node and $j$ denotes other leaf nodes. As there are several paths in a context, we further aggregate the path embeddings by the summation function to produce the \textit{context embedding}, $\mathbf{c_{i}} = \sum_{j \in \mathcal{L}(l_k) \setminus \{i\}}{\mathbf{p_{i\rightarrow j}}}$. On the other hand, the assignments are simply encoded through an \textit{one-hot} encoding. Therefore the final {\bf operand embedding} $\mathbf{x_i} = (\mathbf{c_i} || \mathbf{v_i}),\ \mathbf{x_i} \in \mathbb{R}^{d_c + d_v}$ is the concatenation of the {\bf context embedding} ($\mathbf{c_i} \in \mathbb{R}^{d_c}$), and the corresponding one-hot {\bf assignment encoding} ($\mathbf{v_i} \in \mathbb{R}^{d_v}$).\\


\smallskip

\noindent \bem{Computing weighted sum}:
We propose a novel module to compute attention weights using the \textit{operand embeddings}. It has 
an (i) aggregation layer and an (ii) attention layer.

\smallskip

\noindent i) {\bf Aggregation layer.}
From the domain perspective, we 
design an aggregation step that 
allows for computing a \textit{relative} importance score for each operand. This \textit{importance} score 
reflects the \textit{relative} influence of an 
operand to determine 
an output value, 
when compared 
to all the other operands in the same statement $l_k$.  We create an \textit{updated operand embedding}, $\mathbf{x_i^{*}}$ which captures the \textit{relative importance} and produces a better representation than $\mathbf{x_i}$.

To do so, we augment a single attention head with a special \textit{aggregation layer}. For each statement $l_k$, we aggregate all the operand embeddings using 
the sum 
function. Further, we aid the 
aggregation with a \textit{skip-connection} weighted by a learnable parameter $\epsilon$. 
The \textit{updated operand embeddings} is as follows:  

\vspace{-5mm}
\[
\mathbf{x_i^{*}} = MLP_{\theta_1}\left(\textstyle\sum_{j \in \mathcal{L}(l_k)} \mathbf{x_j} + \epsilon \cdot  \mathbf{x_i}\right), \mathbf{x_i^{*}} \in \mathbb{R}^{d_a}
\]
\vspace{-6mm}





\smallskip

\noindent ii) {\bf Attention layer.}
We use 
the \textit{updated} embeddings to 
compute the 
attention weights. If a 
statement $l_k$ has $N_{i}^{l_k}$ operands, $\mathbf{X}_{l_k}^{*} \in \mathbb{R}^{N_{i}^{l_k} \times d_a}$ is the matrix obtained from the stacking 
all the \textit{updated} operand embeddings. The $\mathbf{X}_{l_k} \in \mathbb{R}^{N_{i}^{l_k} \times (d_c + d_v)}$ is the operand embedding matrix and $\mathbf{A} \in \mathbb{R}^{N_{i}^{l_k} \times d_a}$ is built by repeating the attention vector $\mathbf{a}$ for $N_{i}^{l_k}$ times. Attention weights are computed as follows: 

\vspace{-5mm}

\[
Attention(A, X_{l_k}^{*}, X_{l_k}) = softmax(A X_{l_k}^{*T}) X_{l_k}
\]

\vspace{-2mm}



\smallskip

\noindent \bem{Final prediction:}
We use the 
attention weights to compute a weighted summation of operand embeddings to generate 
the final embedding for a statement $l_k$. We use this embedding 
for the final prediction through $MLP_{\theta_2}$.  
We train $\pname$ 
end-to-end and it \textbf{learns execution semantics} from the simulation data. Attention weights for each prediction contain rich \textit{execution} information which we 
use for \textbf{bug localization}. \\




\smallskip

\noindent {\bf Training Loss:} 
We train our model with a cross-entropy loss using the 
ground-truth 
extracted from the simulation traces. 
To account for an unbalanced data set, we weight our loss with 
{\em inverse class frequencies} in the train set. 
We observe that training with this 
loss function barely updates the value of the attention head from its initialization value. 
To address this, we augment the previous loss function with a \textit{regularization} term 
to update the model parameters more accurately. For a batch $\mathcal{X_B}$ of training samples, the updated loss is 
as follows: 

\vspace{-4mm}

\begin{equation}
\mathcal{L(X_B)} = \frac{\sum_{i=1}^{N} CE(y_i, \tilde{y}_i)}{\sum_{i=1}^{N} w_0 \mathbbm{1}_{\tilde{y}_i = 0} + w_1 \mathbbm{1}_{\tilde{y}_i = 1}} + \frac{\alpha}{N} \sum_{i=0}^{N} \frac{1}{\lVert X_i^{*} \rVert}    \nonumber
\end{equation}


\vspace{-1mm}

\noindent where $y_i = y(l_i), \forall i \in \mathcal{X_B}$ is model prediction, $\tilde{y}_i$ is ground-truth, $\mathbbm{1}$ is the indicator function, $\alpha$ is a weighting factor for the regularization term, ($w_0$,$w_1$) are loss weights, $X_{i}^{*}$ are the \textit{updated} operand embeddings for 
statement $l_i$, and $\lVert \cdot \rVert$ is a matrix 
norm. 
While the {\em first term is to 
train a predictor}, the {\em second term is dependent on our bug localization task}. 
Specifically, the second term helps 
to update the aggregation and projection layer parameters and accurately find the \textbf{importance} of each operand by computing attention scores. 

\subsection{Generating explanations of localization} \label{subsec:explainer}

The 
\textit{Explainer} uses the learned execution semantics 
to produce the final heatmap $\mathbb{H}_t$. 
$\pname$ enables \textbf{finer-grained localization} by mapping $\mathbb{H}_t$ back to the Verilog code. 
We generate the heatmap using the 
attention weights produced by our DL 
model on multiple inference steps. 



\smallskip

\noindent {\bf Attention maps}: We refer 
the attention weights for a statement as its {\em explanation} of suspiciousness of being buggy. Intuitively, 
attention weights capture execution effect of a design statement on the target output. If the attention weights of a statement substantially differ in passing and failing traces, it indicates significantly different execution behavior of the statement, thereby raising suspicion. 
We 
aggregate the explanations of all the statements in the dynamic slice of ${\bf t}$ and create an {\em attention map} which represents relative suspiciousness of different statements in the slice. 



\smallskip

\noindent {\bf Aggregated maps and heatmap generation}: A heatmap can be generated by comparing attention maps generated from \textit{failure} 
traces ($\mathcal{T}_f$) and {\em correct} 
traces ($\mathcal{T}_c$). A trace is a {\em failure} trace if a bug is symptomatized 
at one of the outputs of $\mathbf{D}$, otherwise, it is considered a {\em correct} trace. 
We futher aggregate all the attention maps within the two sets, producing two \textit{aggregated} maps that summarize the design behavior 
in two different sets of traces. 
The aggregation is performed by computing a {\em statement-wise} average of the predicted attention weights. We denote these aggregation as $\mathbb{F}_t$ and $\mathbb{C}_t$ for failing and correct traces, 
respectively. To generate a heatmap $\mathbb{H}_t$, we make a {\em statement-wise} comparison of these two maps. Intuitively, if a design statement shows a significantly different execution behavior in the correct and failure traces, then the corresponding attention weights for that statement would be significantly different in 
$\mathbb{F}_t$ and $\mathbb{C}_t$. {\bf We highlight those design statement(s) as likely buggy candidates for debugging}. 


Depending on the construction of 
the dynamic slices for 
${\bf t}$, $\mathbb{F}_t$ and $\mathbb{C}_t$ may not contain information about the same set of statements. 
Thus, for a systematic comparison, we identify following three possible scenarios: 


\begin{itemize}
    \item \textbf{A statement $l_k$ is present in $\mathbb{C}_t$ but absent in $\mathbb{F}_t$}. We can infer that $l_k$ is \bem{not suspicious}, as 
    ${\cal T}_f$ do not record executions for $l_k$,
    thus it cannot cause a bug to symptomatize at ${\bf t}$.  
    
    \item \textbf{A statement $l_k$ is present in $\mathbb{F}_t$ but absent in $\mathbb{C}_t$}. The execution of $l_k$ affects the 
    $\mathbf{t}$ only in ${\cal T}_f$, 
    causing a bug to symptomatize at ${\bf t}$. 
    Thus, we mark $l_k$ 
    as \bem{suspicious} and copy its attention weights from the $\mathbb{F}_t$ to the final heatmap $\mathbb{H}_t$.
    
    \item \textbf{A statement is present in 
    $\mathbb{F}_t$ and $\mathbb{C}_t$}. This is a  non-trivial case requiring further analysis. 
    We identify this statement as \bem{highly likely suspicious} if its attention weights \textbf{significantly differ} between $\mathbb{F}_t$ 
    and $\mathbb{C}_t$. 
    To compute this difference, we use a \textit{normalized} norm-1 distance. We apply min-max normalization on this distance, with $min=0$ and $max=2$, as a norm-1 distance between two sets of our attention weights always falls in this range. 
    The computed difference is compared with a user-defined \textit{suspicious threshold}; we set it to $0.10$ for our analysis. When a statement is classified as \textit{suspicious}, we move $\mathbb{F}_t$ weights for $l_k$ 
    to the final $\mathbb{H}_t$.
\end{itemize}

Following this process, the final heatmap $\mathbb{H}_t$ contains only attention weights of 
\textit{candidate buggy statements}. 

\section{Experimental Setup}
\label{sec:exp_setup}




\noindent {\bf Dataset generation}:
We train $\pname$ on a \textit{synthetic} data set \addt{to force variability in training data, necessary to achieve generalization}. Our synthetic {\em Random Verilog Design Generator} (RVDG) randomly generates Verilog designs using pre-defined template as follows. 
It has two distinct parts. In the first part, we have a {\em clocked always} block ($\mathcal{C}$) that acts as a memory element to remember the current design state. The other {\em non-clocked always} block ($\mathcal{NC}$) computes the next state of the design based on the current state of the design and current inputs. It also assigns Boolean values to one or more design outputs. The $\mathcal{NC}$ contains multiple {\em if-else-if} blocks, each block computing using blocking assignments. RVDG randomly generates legal blocking assignments following Verilog's grammar. RVDG ensures the presence of interdependencies among design variables to create data flows. RVDG also controls the maximum number of operands and Boolean operators in each design statement.

\smallskip

\noindent {\bf Training model}: 
We used the \textit{Adam}~optimizer\cite{kingma2017adam} for \textit{mini-batch} stochastic gradient descent, with learning rate $lr = 10^{-3}$ and weight decay $wd = 10^{-5}$. We create batches 
by sampling statements and their associated input vectors from the training set. In the case of statement \textit{inter-dependencies} 
requiring past predictions, ground truth results are used during training. At inference time, we obtain predictions following the correct statement order. This training strategy allows much faster training. We also set $d_a = 32$ as size for the attention vector, and $d_c = 16$ as the dimension of the context embedding vector. With this configuration, $\pname$ takes a few minutes to train.



\smallskip

\noindent {\bf Bug injection}: To validate the bug localization effectiveness 
we inject bugs of varying complexity  
on real-designs. 
\Cref{tab:module_desc} shows details for the realistic designs of our analysis. In this setup, \textit{data-centric} bugs are introduced by automatically \textit{mutating} the designs according to a pre-defined set of mutation rules.  
Specifically, we introduce \textit{three} different types of bugs: (i) \bem{Negation}, where the injected bug introduces 
a wrong ``NOT" ($\neg$) in front of an operand or removes an existing one; (ii) \bem{Variable misuse}, where a variable name is changed with another one, possibly syntactically similar. This replicates traditional human {\em copy-paste} errors; (iii) \bem{Operation substitution}, where a Boolean operator is substituted with a wrong one, \eg, an ``OR'' ($\vee$) replaced with a ``AND'' ($\wedge$). 
To avoid undesirable masking behaviors, 
we inject one bug per mutated design. 
A bug is {\em observable} if the bug symptomatizes at 
design outputs. 

\smallskip

\noindent {\bf Target predictor selection}: 
Predictor performance could \textit{theoretically} rely on $\alpha$, the weighing factor of the \textit{regularization} term introduced for the bug localization task (\cf, Training loss in~\Cref{sec:dl_model}). Hence, we evaluate predictor performance for different $\alpha$ values.
We evaluate the predictors based on the quality of predicting the output value of a statement, on a test-set of \textit{holdout} synthetic designs. We assume that the predictors with higher prediction accuracy are 
able to compute better statement representations 
and in general 
learn better \textit{features} related to execution semantics. As we have an unbalanced test set, we also compute \textit{precision} and \textit{recall} obtained for both target bit values. We report these results in~\Cref{tab:predictor_results}. We 
observe that there is no substantial change in performances. 
However, $\alpha=0.10$ produces slightly higher prediction accuracy, 
precision, and recall. Hence, we choose $\alpha = 0.10$ for our experiments.


\begin{table}
\vspace{-2mm}
\centering
\caption{
\label{tab:module_desc}
{\bf Details of modules in our \textit{localization} test set}. 
}
\vspace{-2mm}
\begin{tabular}{|l|c|l|}
\hline
\multicolumn{1}{|c}{Module Name} & \multicolumn{1}{|c}{Line of Codes} & \multicolumn{1}{|c|}{Short Description} \\
\hline
wb\_mux\_2           & 65                         & \textit{\begin{tabular}[c]{@{}l@{}}Wishbone 2-port \\ Multiplexer\end{tabular}} \\ \hline
usbf\_pl             & 287                        & USB2.0 Protocol Layer                                                           \\ \hline
usbf\_idma           & 627                      & \begin{tabular}[c]{@{}l@{}}USB2.0 Internal \\ DMA Controller\end{tabular}       \\ \hline
ibex\_controller     & 459                        & \begin{tabular}[c]{@{}l@{}}Ibex RISC-V \\ Processor Controller\end{tabular}     \\ \hline
\end{tabular}
\vspace{-2mm}
\end{table}

\begin{table}
\centering
\caption{
\label{tab:predictor_results}
{\bf Results on test-set obtained for different weighting $\alpha$ factors.} Pr = Precision, Re = Recall, Target = bit-level prediction class. 
We choose the predictor with $alpha = 0.1$ as it exhibits the best execution semantics learning abilities, correctly predicting 
bit values of the outputs in almost all cases.
}
\vspace{-1mm}
\begin{tabular}{|l|c|c|c|}
\hline
\textbf{$alpha$} & \multicolumn{1}{l|}{\textbf{Acc. (\%)}} & \multicolumn{1}{l|}{\textbf{Pr/Re (Target 0)}} & \multicolumn{1}{l|}{\textbf{Pr/Re (Target 1)}} \\ 
\hline 
0.01 & 96.5 & 0.99/0.96  & 0.93/0.98  \\ 
\hline
0.05 & 93.8 & 0.96/0.94 & 0.90/0.93 \\ 
\hline
0.10 & \textbf{98.0} & \textbf{0.98/0.99}  & \textbf{0.98/0.96} \\ 
\hline
0.15 & 95.6 & 0.95/0.97 & 0.96/0.92 \\ 
\hline
0.20 & 96.7 & 0.97/0.98 & 0.96/0.95 \\ 
\hline
0.25 & 97.7 & 0.97/0.99 & \textbf{0.99/0.95} \\ 
\hline
\end{tabular}
\vspace{-6mm}
\end{table}

\begin{table*}
\vspace{-2mm}
\centering
\caption{
\label{tab:bug_results}
{\bf Bug coverage 
for bug-localization on realistic 
designs.} For each injected bug-type, we show 
number of design versions. 
We compute 
top-1 coverage 
by considering 
observable bugs on the target output and the number of localized bugs.
}
\vspace{-1mm}
\begin{tabular}{|l|l|c|c|c|c|c|}
\hline
\multicolumn{1}{|c}{Design Name} &
\multicolumn{1}{|c}{Target} & 
\multicolumn{4}{|c}{Injected Bugs} &
\multicolumn{1}{|c|}{$\uparrow$ top-1 Coverage (\%)} \\
\cline{3-6}
& & \textit{Negation} & \textit{Operation} & \textit{Misuse} & Total (Observable) & \\
\hline
wb\_mux\_2       & wbs0\_we\_o            & 2                  & 2                   & 4                & 8 (8)           &   87.5\% (7/8)\\ \hline
wb\_mux\_2       & wbs0\_stb\_o           & 2                  & 2                   & 4                & 8 (8)           &   87.5\%  (7/8) \\ \hline
\hline
\textbf{wb\_mux\_2} & - & 4 & 4 & 8 & 16 (16) & \textbf{87.5\% (14/16)} \\ \hline
\hline
usbf\_pl         & match\_o               & 5                  & 8                   & 9                & 22 (10)         &   60.0\% (6/10) \\ \hline
usbf\_pl         & frame\_no\_we          & 3                  & 4                   & 9                & 16 (12)         &   66.6\% (8/12) \\ \hline
\hline
\textbf{usbf\_pl} & - & 8 & 12 & 18 & 38 (22) & \textbf{63.6\% (14/22)} \\ \hline
\hline
usbf\_idma       & mreq                   & 3                  & 4                   & 6                & 13 (12)         &   50.0\% (6/12) \\ \hline
usbf\_idma       & adr\_incw              & 2                  & 2                   & 8                & 12 (12)         &   91.6\% (11/12) \\ \hline
\hline
\textbf{usbf\_idma} & - & 5 & 6 & 14 & 25 (24) & \textbf{70.8\% (17/24)} \\ \hline
\hline
ibex\_controller & stall                  & 4                  & 6                   & 12               & 22 (22)         &   95.4\% (21/22) \\ \hline
ibex\_controller & instr\_valid\_clear\_o & 3                  & 4                   & 12               & 19 (19)         &   100\% (19/19)            \\ \hline
\hline
\textbf{ibex\_controller} & - & 7 & 10 & 24 & 41 (41) & \textbf{97.6\% (40/41)} \\
\hline
\hline
\textbf{Overall} & - & \textbf{24} & \textbf{32} & \textbf{64} & \textbf{120 (103)} & \textbf{82.5\% (85/103)} \\
\hline
\end{tabular}
\end{table*}


\begin{figure*}
    \centering
    \includegraphics[scale=0.60]{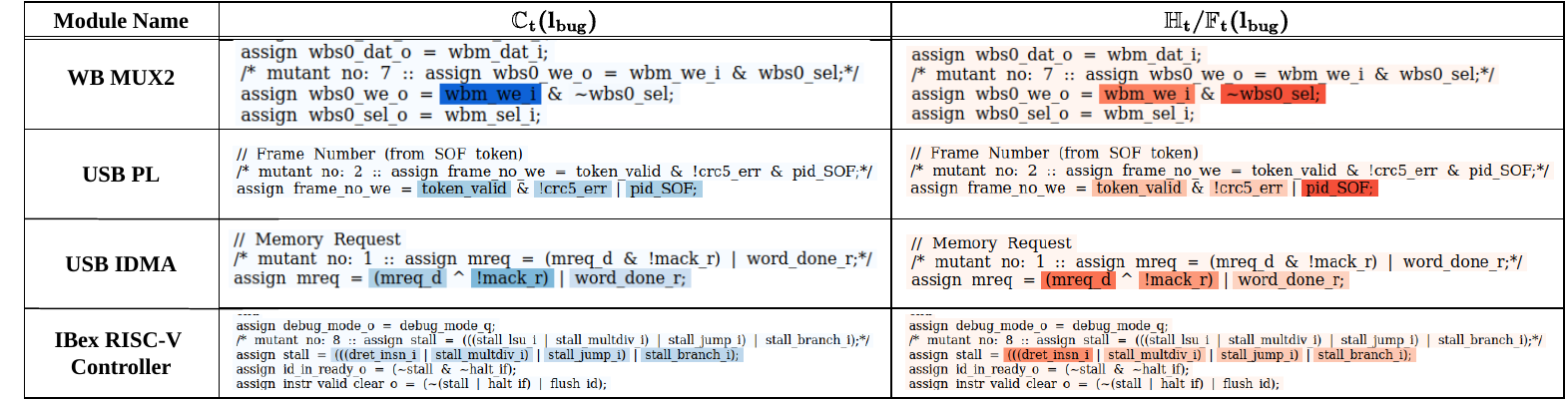}
    \caption{{\bf$\pname$ qualitative results on realistic 
    designs: examples of $\pname$ generated heatmaps}. For comparison, we also report the operand importance scores (blue, deeper is more important) in $\mathbb{C}_t$, against which important scores (red, deeper is more important) in $\mathbb{F}_t$ are compared. 
    The rightmost column reports the \textit{suspiciousness} score for the statement $l_{bug}$ ($\ie$, the statement with the root cause).
    Note that $\mathbb{H}_t(l_{bug}) = \mathbb{F}_t(l_{bug}) $ when \textit{suspiciousness} score for $l_{bug}$ is higher than threshold.
    \label{fig:qualitative_results}}
    \vspace{-4mm}
\end{figure*}

\section{Experimental Results}
\label{sec:exp_results}

\subsection{Bug Coverage Results on Open-Source Designs}
\label{subsec:quantitative}

In this section, we evaluate the effectiveness of $\pname$ for bug localization using realistic designs. We show our localization results in~\Cref{tab:bug_results}.
%
We compute a bug coverage metric for each design-target pair as the ratio between the number of localized bugs and the total number of observable bugs. We consider a bug as 
localized when the highest suspiciousness score in the heatmap  $\mathbb{H}_t$ is assigned to the statement containing the root cause. Thus, we compute a top-1 bug coverage metric. Note that a localization error (\ie, coverage reduction) could be due to either the suspiciousness score is not the highest in heatmap or a statement's exclusion from the heatmap due to suspiciousness score lower than the user-defined threshold. 
We consider the highest suspiciousness scores 
after running the same $\pname$ instance over multiple simulation runs. 

We observe that $\pname$ is effective in localizing different bug types in data-flow, especially for the \textit{ibex\_controller} ($97.6\%$ of top-1 bug coverage) and the \textit{wb\_mux\_2} ($87.5\%$ of top-1 bug coverage) designs.
The two USB modules, \textit{usbf\_idma} and the \textit{usbf\_pl} have lower bug coverage with $70.8\%$ and $63.6\%$ coverage respectively. For the former, this is due to the difficulty in finding the injected bug on the target output as we 
report in \Cref{tab:bug_results}.  Considering all designs and injected bugs, we obtain an overall top-1 bug localization coverage of $82.5 \%$. This experiment shows that $\pname$ is a promising approach for precise bug localization 
of data-centric bugs. Note, we train $\pname$ on synthetic designs, yet it localizes well in realistic designs. This empirical evidence shows that the learned knowledge is transferable making our approach as a promising one for broader adoption.


\subsection{Exact Root-Cause Localization via $\pname$ Heatmaps}
\label{subsec:qualitative}

$\pname$ allows for further localization via mapping importance scores stored in heatmap $\mathbb{H}_t$ back to the RTL code. In this experiment we evaluate how these importance scores help in bug localization. 
We show some examples of $\pname$ generated heatmaps on realistic designs in~\Cref{fig:qualitative_results}. Together with heatmaps $\mathbb{H}_t$, we also provide a visualization of scores stored in $\mathbb{C}_t$. 
To offer better visualization, we discretize the range $[0, 1]$ into 
bins, 
assigning a different color intensity to each bin, in the scale of reds for $\mathbb{H}_t$ and blues for $\mathbb{C}_t$. We also report the corresponding \textit{suspiciousness} score ${\bf d}(\mathbb{F}_t(l_{bug}), \mathbb{C}_t(l_{bug}))$. 

Heatmaps visualized in~\Cref{fig:qualitative_results} show how, in failure traces ${\cal T}_f$, an \textbf{higher importance score is assigned to operands that are source of bugs or are directly involved in an buggy operation}. We also show how \textbf{different importance scores are assigned to the same operands for correct traces} ${\cal T}_c$, where the \textbf{bug is instead masked}.  
Results obtained for the \textit{usbf\_idma} and \textit{ibex\_controller} modules also justify our choice of selecting $0.1$ as our default threshold for the suspiciousness score, as a score higher than this corresponds to a visual difference in the two maps. This experiment shows that
$\pname$ is able to 
learn low-level execution semantics details. It can effectively use such learned knowledge to localize the bug and provide necessary explanations at source-code level for debugging.


\section{
Conclusion}
\label{sec:rel_work}

We present $\pname$, which relies on Attention 
and LSTM-based DL model to learn RTL execution semantics. $\pname$ uses a novel approach to repurpose the learned semantic knowledge for bug localization. $\pname$ is unique as compared to the other state-of-the-art techniques as it localizes the bug and generates explanation automatically to make its decision accessible to a debugging engineer. We also show $\pname$ is particularly effective for data-flow bug localization and its learned knowledge is transferable. Albeit very preliminary, our approach shows promising results and it is compatible with current verification flows without any additional artifacts. 


\scriptsize


\end{document}